# Topological Insulator State due to Finite Spin-Orbit Interaction in an Organic Dirac Fermion System

Toshihito Osada*

*Institute for Solid State Physics, University of Tokyo,*

*5-1-5 Kashiwanoha, Kashiwa, Chiba 277-8581, Japan.*

Recently, Valenti et al. pointed out the significance of the spin-orbit interaction (SOI) in organic materials, and explained the anomalous insulating behavior of a Dirac semimetal $\alpha$-(BEDT-TTF)$_2$I$_3$ at low temperatures in terms of the small SOI gap. We propose a lattice model with plausible SOI coupling, and indicate that the gapped state is a topological insulator. This model is an organic analogue of the Kane-Mele model for graphene.

An organic conductor $\alpha$-(BEDT-TTF)$_2$I$_3$ has attracted a great deal of attention because of its two-dimensional (2D) massless Dirac fermion state (Dirac semimetal) under high pressure [1]. This fact was originally pointed out theoretically [2], and later confirmed by various experiments, such as magnetotranport [3, 4], specific heat [5], and NMR [6]. One of the unsolved subjects has been the anomalous resistance increase at low temperatures in the Dirac semimetal state [7, 8]. The insulating temperature dependence seems to follow the logarithmic rule rather than the Arrhenius rule [7].

Recently, Valenti et al. pointed out the significance of spin-orbit interaction (SOI) in organic conductors based on the first principles calculation [9]. They estimated SOI on a BEDT-TTF layer to be in the order of 1 ~ 2 meV. This value is small but possible to occur observable effects. They ascribed the insulating behavior in $\alpha$-(BEDT-TTF)$_2$I$_3$ to the energy gap opening at the Dirac point due to SOI.

In this paper, we discuss the gapped state in $\alpha$-(BEDT-TTF)$_2$I$_3$ under finite SOI. We illustrate that a plausible SOI opens a gap in the Dirac cone, and the gapped state must be a topological insulator. The present model is an organic version of the Kane-Mele model for graphene [10]. In contrast to the Kane-Mele model, which assumed unrealistic next nearest neighbor hopping in the honeycomb lattice, we introduce no additional hopping in the anisotropic triangular lattice of $\alpha$-(BEDT-TTF)$_2$I$_3$. It is similar to the other model [11], which was an organic version of the Haldane model [12].

Generally, SOI is proportional to $\{\mathbf{p} \times \nabla U(\mathbf{r})\} \cdot \boldsymbol{\sigma}$, where $\mathbf{p}$, $\nabla U(\mathbf{r})$, and $\boldsymbol{\sigma}$ are the electron momentum, potential gradient, and electron spin, respectively. Since the electron orbital motion is restricted on the two-dimensional (2D) BEDT-TTF layer schematically shown in Fig. 1(a), $\mathbf{p}$ and $\mathbf{r}$ are considered to be on the 2D plane in the effective mass approximation, so that the effective magnetic field $\mathbf{p} \times \nabla U(\mathbf{r})$ is normal to the 2D plane. Since the normal component $s_z$ of the electron spin $\boldsymbol{\sigma}$ becomes a good quantum number, the Hamiltonian $H(\mathbf{k})$ is decouples into two spin sectors $H(\mathbf{k}, s_z)$ with $s_z = \pm 1$ as the following.

$$H(\mathbf{k}) = \begin{pmatrix} H(\mathbf{k}, +1) & 0 \\ 0 & H(\mathbf{k}, -1) \end{pmatrix}. \tag{1}$$

Here, $H(\mathbf{k}, s_z)$ is a $4 \times 4$ Hamiltonian matrix, of which bases are the Bloch sums constructed from HOMOs of molecules A, A', B, and C with a spin $s_z$.

$$H(\mathbf{k}, s_z) = \begin{pmatrix} \varepsilon_0 & H_{AA'}(\mathbf{k}) & H_{AB}(\mathbf{k}, s_z) & H_{AC}(\mathbf{k}, s_z) \\ H_{AA'}(\mathbf{k})^* & \varepsilon_0 & H_{A'B}(\mathbf{k}, s_z) & H_{A'C}(\mathbf{k}, s_z) \\ H_{AB}(\mathbf{k}, s_z)^* & H_{A'B}(\mathbf{k}, s_z)^* & \varepsilon_0 & H_{BC}(\mathbf{k}) \\ H_{AC}(\mathbf{k}, s_z)^* & H_{A'C}(\mathbf{k}, s_z)^* & H_{BC}(\mathbf{k})^* & \varepsilon_0 \end{pmatrix}, \tag{2}$$

$$H_{AA'}(\mathbf{k}) = a_2 e^{+i\mathbf{k}\cdot\boldsymbol{\tau}_1} + a_3 e^{-i\mathbf{k}\cdot\boldsymbol{\tau}_1},$$

$$H_{BC}(\mathbf{k}) = a_1 e^{+i\mathbf{k}\cdot\boldsymbol{\tau}_1} + a_1 e^{-i\mathbf{k}\cdot\boldsymbol{\tau}_1},$$

$$H_{AB}(\mathbf{k}, s_z) = b_2(1 + i\lambda s_z) e^{+i\mathbf{k}\cdot\boldsymbol{\tau}_2} + b_3(1 - i\lambda s_z) e^{-i\mathbf{k}\cdot\boldsymbol{\tau}_3},$$

$$H_{AC}(\mathbf{k}, s_z) = b_1(1 + i\lambda s_z) e^{+i\mathbf{k}\cdot\boldsymbol{\tau}_3} + b_4(1 - i\lambda s_z) e^{-i\mathbf{k}\cdot\boldsymbol{\tau}_2},$$

$$H_{A'B}(\mathbf{k}, s_z) = b_2(1 + i\lambda s_z) e^{-i\mathbf{k}\cdot\boldsymbol{\tau}_2} + b_3(1 - i\lambda s_z) e^{+i\mathbf{k}\cdot\boldsymbol{\tau}_3},$$

$$H_{A'C}(\mathbf{k}, s_z) = b_1(1 + i\lambda s_z) e^{-i\mathbf{k}\cdot\boldsymbol{\tau}_3} + b_4(1 - i\lambda s_z) e^{+i\mathbf{k}\cdot\boldsymbol{\tau}_2}.$$

The above matrix is basically the conventional tight-binding model for $\alpha$-(BEDT-TTF)$_2$I$_3$ except spin dependent terms representing SOI [1, 11]. $\mathbf{k} = (k_x, k_y)$ is a 2D wave vector, and $\boldsymbol{\tau}_1 = (0, a/2)$, $\boldsymbol{\tau}_2 = (b/2, -a/4)$, and $\boldsymbol{\tau}_3 = (b/2, a/4)$, where $b$ and $a$ are lattice constants along the $x$ and $y$ directions, respectively. The energy of the HOMO, $\varepsilon_0$, is taken as zero. Transfer integrals are chosen as $a_1 = -0.038$ eV, $a_2 = +0.080$ eV, $a_3 = -0.018$ eV, $b_1 = +0.123$ eV, $b_2 = +0.146$ eV, $b_3 = -0.070$ eV, and $b_4 = -0.025$ eV, so as to give a Dirac semimetal in the case that SOI strength, $\lambda$, is zero. These parameters correspond to the case of uniaxial pressure, $P_a = 0.4$ GPa, in previous calculations [1].

In the tight-binding model, SOI adds the spin-dependent complex correction to transfer integrals. The intra-chain transfer integrals, $a_1$, $a_2$, and $a_3$, have no SOI correction, since the average potential gradient $\nabla U(\mathbf{r})$ is orthogonal to the hopping direction (//$\mathbf{p}$) along the AA' or BC molecular chain (//$y$-axis), due to mirror symmetry with respect to the chain, ignoring molecular shape. On the other hand, the inter-chain transfer integrals, $b_1$, $b_2$, $b_3$, and $b_4$, have finite SOI correction, since $\mathbf{p} \times \nabla U(\mathbf{r})$ becomes finite due to potential asymmetry with respect to the inter-chain hopping direction. For simplicity, we assume that the correction term to $b_i$ ($i = 1, 2, 3, 4$) takes the form of $i\lambda b_i \nu_i s_z$, where $\lambda$ is a common parameter indicating SOI strength, and $\nu_i$ takes the value of $\pm 1$ depending on the asymmetry of both sides of its hopping path.

In actual $\alpha$-(BEDT-TTF)$_2$I$_3$, it was experimentally clarified that the electron density on each molecule is not uniform (charge disproportionation) [13]. The molecular charge on each molecule follows $B > A = A' > C$, as shown by the size of clouds in Fig. 1(a). In the above model, the SOI sign $\nu_i$ was determined considering $\nabla U(\mathbf{r})$ due to the charge disproportionation, although the difference of molecular charge is too small to cause observable SOI.

The above model gives the band dispersion with a finite energy gap between the third (valence) and fourth (conduction) bands as shown in Fig. 1(b). In below, we see that this insulating state is a topological insulator. The lattice structure of $\alpha$-(BEDT-TTF)$_2$I$_3$ has space inversion symmetry with an inversion center at the midpoint of A and A' molecules. In each band, up-spin ($s_z = +1$) and down-spin ($s_z = -1$) subbands are degenerated under the inversion symmetry. Fu and Kane proposed a simple method to judge whether an insulator with inversion symmetry is a topological insulator or not [14]. In the 2D Brillouin zone of $\alpha$-(BEDT-TTF)$_2$I$_3$, there exist four time-reversal invariant wave numbers (TRIMs), which satisfy $-\mathbf{k}_{TRIM} = \mathbf{k}_{TRIM} + \mathbf{G}$ with a reciprocal lattice vector $\mathbf{G}$, $\mathbf{k}_{TRIM} = (0,0)$, $(\pi/b, 0)$, $(0, \pi/a)$, and $(\pi/b, \pi/a)$, which are referred as Γ, X, Y, and M, respectively, in Fig. 1(b). Here, let us focus on parity $P_n(\mathbf{k}_{TRIM})$, which is the eigenvalue (+1 or −1) of the space inversion operator, at $\mathbf{k}_{TRIM}$ for the $n$-th spin-degenerated band. According to Fu and Kane, the condition for a topological insulator with inversion symmetry is that the product of $P_n(\mathbf{k}_{TRIM})$ for all $\mathbf{k}_{TRIM}$ and all $n$ of occupied bands is equal to −1. In Fig. 1(b), $P_n(\mathbf{k}_{TRIM})$ are indicated by the symbols, "+" or "−", corresponding to $P_n(\mathbf{k}_{TRIM}) = +1$ or −1, respectively. Since the parity product for three occupied bands below the Fermi level is equal to −1, we can conclude that the present model describes a non-trivial topological insulator.

At the zero SOI limit ($\lambda \to 0$), the present model represents a gapless Dirac semimetal with a pair of Dirac cones. Piechon and Suzumura discussed that the Dirac semimetal state in $\alpha$-(BEDT-TTF)$_2$I$_3$ is stable as long as the Fu-Kane parity condition is satisfied [15]. This means that $\alpha$-(BEDT-TTF)$_2$I$_3$ with SOI gap is a topological insulator, as long as SOI does not cause any band inversion accompanied by parity change.

In fact, the present model gives the Berry curvature indicating to be a topological insulator. Since $H(\mathbf{k})$ is decoupled into two spin sectors, $H(\mathbf{k}, +1)$ and $H(\mathbf{k}, -1)$, we can calculate Berry curvature separately for each spin-subband [11]. Figure 2(a) shows the $z$-component of Berry curvature of the spin subbands of the valence and conduction bands, $E_{3\uparrow}(\mathbf{k})$, $E_{3\downarrow}(\mathbf{k})$, $E_{4\uparrow}(\mathbf{k})$, and $E_{4\downarrow}(\mathbf{k})$. Each Berry curvature exhibits two peak structures around two valleys with the same sign. However, the sign of Berry curvature is opposite for up-spin and down-spin subbands. This means that the carriers with different spin obtain opposite directions of anomalous velocity under the external in-plane electric field. This causes no charge current but finite spin current in the direction perpendicular to the electric field, namely, the spin Hall effect. Moreover, according to the TKNN theory [16, 17], the contribution of the all occupied up-spin or down-spin subbands to the off-diagonal conductivity is quantized to $-N_{\text{spin}}e^2/h$, where $N_{\text{spin}}$ is an integer called the spin-Chern number. Therefore, the system shows the quantum spin Hall effect.

Next, we investigate the edge state of a 2D nanoribbon of $\alpha$-(BEDT-TTF)$_2$I$_3$ parallel to the $y$-axis, which has two types of edges, the AA'-edge and BC-edge [11, 18]. The calculated $k_y$-dispersion of the energy spectrum is shown in Fig. 2(b). The isolated branches which appear in the bulk gap are the edge states localized around the AA'-edge or BC-edge as indicated by labels. We can see that a pair of edge states with opposite spin and group velocity at the Fermi level, called the helical edge state, appear around the AA'-edge and BC-edge. The helical edge state carries no charge current but finite spin current along the sample edge. The appearance of the metallic helical edge state is one of the most characteristic features of topological insulators.

In conclusion, we have constructed a model for an organic Dirac fermion system with finite SOI, which opens a gap at the Dirac point. We have shown that it is a topological insulator state with the helical edge state, as long as SOI does not cause any band inversion. Therefore, if SOI opens a significant gap causing resistance increase observed at low temperature in $\alpha$-(BEDT-TTF)$_2$I$_3$, the resistance increase must be limited by the helical edge transport. Conversely, if the resistance increase shows no saturation at low temperature limit, we can conclude that SOI is not necessarily significant in

$\alpha$-(BEDT-TTF)$_2$I$_3$.

**Acknowledgements**

The author thanks Mr. K. Yoshimura, Ms. A. Mori, and Dr. M. Sato for valuable discussions. He also thanks Prof. N. Tajima and Dr. T. Miyazaki for their useful comments. This work was supported by JSPS KAKENHI Grant Numbers JP25107003 and JP16H03999.

**Figure 1** (Osada)

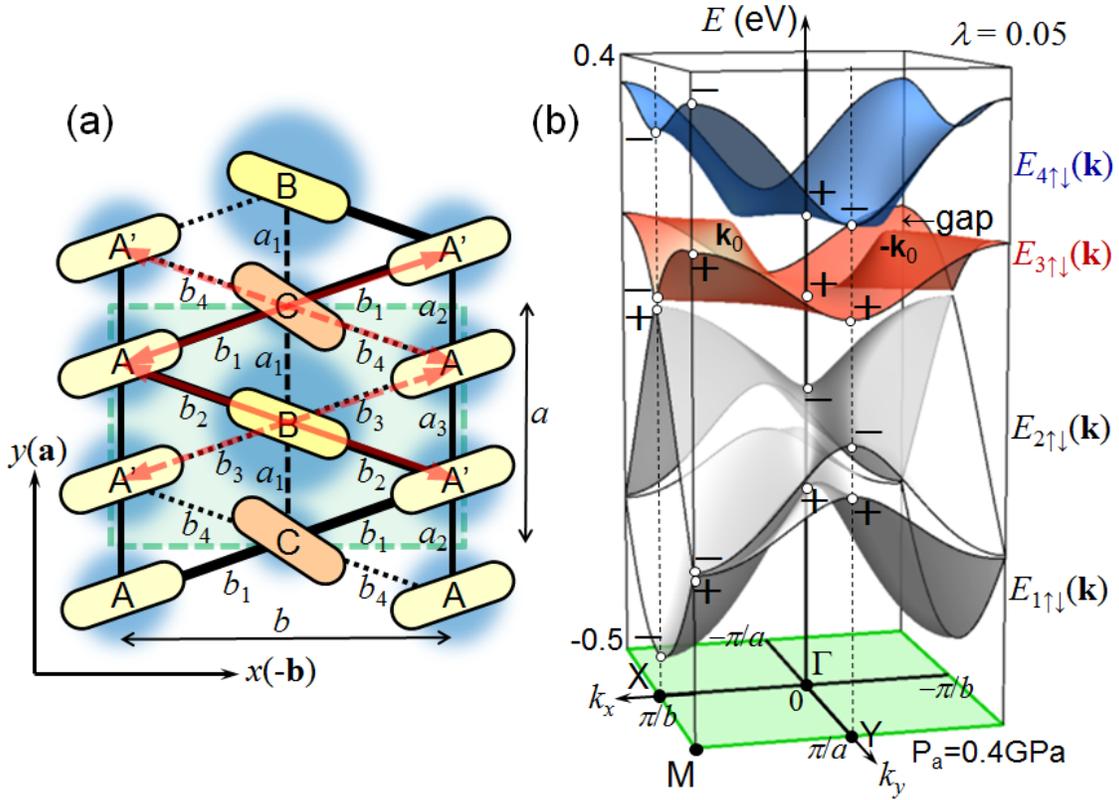

**FIG. 1.** (color online)

(a) Schematic of lattice structure of a conducting layer of α-(BEDT-TTF)$_2$I$_3$. Transfer integrals are shown. The pale solid and dashed arrows indicate the hopping with diferent sign of SOI. The charge disproportionation is represented by the size of clouds. (b) Band structure of the α-(BEDT-TTF)$_2$I$_3$ under finite SOI ($\lambda$ = 0.05). The parity of each spin-degenerated band at TRIMs is indicated by "+" or "−".

**Figure 2** (Osada)

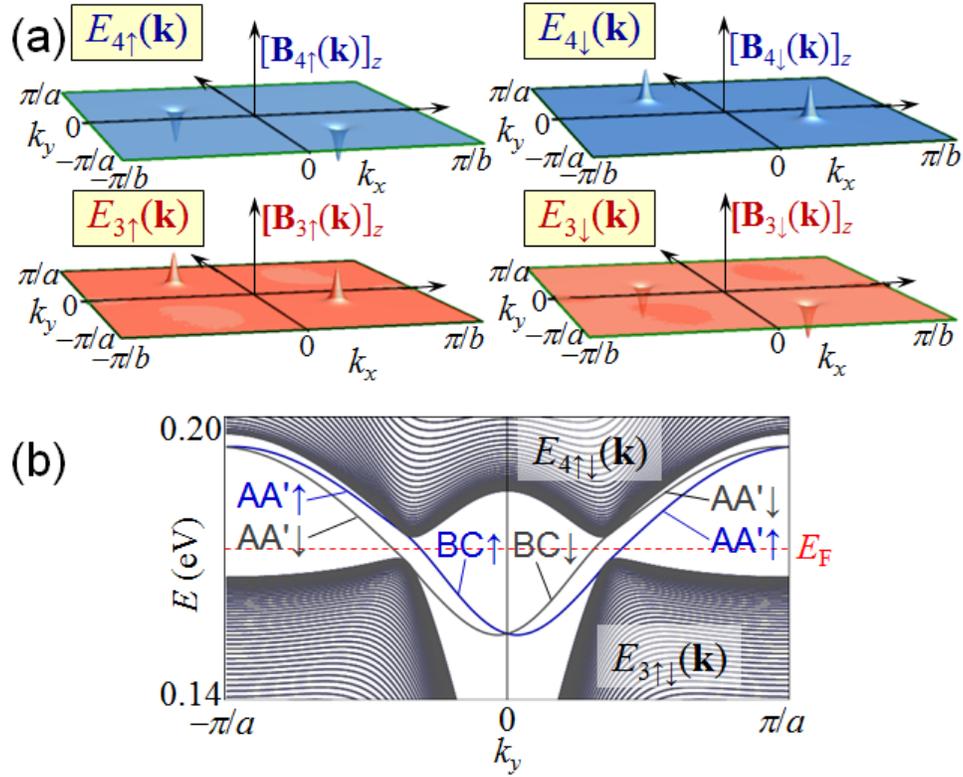

**FIG. 2.** (color online)

(a) Berry curvature of spin-subbands of the valence and conduction bands for $\lambda = 0.05$.

(b) Energy dispersion of the nanoribbon of $\alpha$-(BEDT-TTF)$_2$I$_3$ parallel to the **a**-axis for $\lambda = 0.05$. The helical edge state appears along both the AA'- and BC-edges.